\documentclass[aps,prd,reprint,showpacs,amsmath,amssymb,superscriptaddress,floatfix]{revtex4-1}
\usepackage{epsf}
\usepackage{slashed}
\usepackage{ulem}
\usepackage{cancel}
\usepackage{color,bm}
\usepackage[colorinlistoftodos]{todonotes}
\usepackage{graphicx}
\usepackage{amsfonts}
\usepackage{amssymb}
\usepackage{rotating}
\usepackage{booktabs}
\usepackage{xcolor}
\usepackage{soul}

\usepackage[colorlinks=true,citecolor=cyan,urlcolor=blue,bookmarks=true,bookmarks=true,bookmarksopen=true,bookmarksnumbered=true,bookmarksopenlevel=3]{hyperref}

\definecolor{airforceblue}{rgb}{0.36, 0.54, 0.66}
\definecolor{steelblue}{rgb}{0.27, 0.51, 0.71}
\definecolor{amber}{rgb}{1.0, 0.49, 0.0}

\def\comment#1{}

\usepackage{graphicx}
\begin{document}

\title{Top FCNC interactions through dimension six four-fermion operators at the electron proton collider}
\author{\textsc{Wei Liu}}
\email{wei.liu.16@ucl.ac.uk}
\affiliation{University College London, Gower Street, London WC1E 6BT, UK}
\author{\textsc{Hao Sun}}
\email{haosun@mail.ustc.edu.cn\ \ haosun@dlut.edu.cn}
\affiliation{Institute of Theoretical Physics, School of Physics, Dalian University of Technology, \\ No.2 Linggong Road, Dalian, Liaoning, 116024, P.R.China}

\date{\today}

\vspace{1cm}

\begin{abstract}
We investigate the flavour changing neutral currents (FCNC) generated by dimension six four-fermion operators at the Large Hadron-Electron Collider (LHeC) proposed at CERN in an effective approach. This is performed by monte carlo analysis at the full detector level, as the background is successfully reduced by using invariant mass scheme to require the final states to reconstruct the mass (transverse mass) of the top quark (W boson). Our analysis shows that the future electron proton colliders like the LHeC can probe competitive limits for the top FCNC dimension six four fermion operators such as $C_{lq}^{(1)ee31} < 0.0647$, $C_{lu}^{ee31} < 0.109$, $C_{lequ}^{(1)ee31} < 0.217$ and $C_{lequ}^{(3)ee31} < 0.0209$ in the Warsaw basis.
\vspace{3cm}
\end{abstract}

\maketitle
\setcounter{footnote}{0}

\vspace{1cm}

\section{Introduction}
\label{sec:intro}

Although the Standard Model (SM) has been tested in great precision, it is commonly to be considered as an effective field theory which is applicable up to a certain scale $\Lambda$. Following this, several studies \cite{Zhang:2017mls, Buckley:2015lku, Buckley:2015nca, Zhang:2010dr, AguilarSaavedra:2018nen,Chala:2018agk,Hartland:2019bjb} have been made on the top quark, the most important heaviest known particle up to the electroweak scale, based on the dimension six operators \cite{Grzadkowski:2010es,Dedes:2017zog,Dedes:2019uzs} of the Standard Model Effective Field Theory (SMEFT) at the LHC. Among these operators, the ones which introduce flavour changing neutral currents (FCNC) are very sensitive to new physics as they are extremely suppressed by Glashow-Iliopoulos-Maiani (G.I.M) mechanism \cite{Glashow:1970gm} in the SM. Meanwhile, a proposed electron proton collider calling the Large Hadron-Electron Collider (LHeC) \cite{Bruening:2013bga, AbelleiraFernandez:2012cc} is served as one of the possible options at the post-LHC era. As the incoming protons are produced from the LHC beam, the high energy makes it suitable for studies of the top quark. Moreover, the asymmetric beams make LHeC much more suitable for probing top FCNC operators including leptons and quarks in particular. Such dimension six operators can include gauge bosons, while as it introduces much more complex processes and backgrounds, we focus on the operators including four fermions at the LHeC, such as the ones which can produce $e^{-} p \rightarrow e^{-} t$ process.

As the hints of lepton flavour universality violation $b\to s\ell\ell$ and $b\to c\tau\nu$  \cite{Capdevila:2017bsm, Altmannshofer:2014rta, Ghosh:2014awa} at the LHCb raised a lot attention, the limits for our interests are mainly reinterpreted through the CKM matrix \cite{Greljo:2017vvb,Jung:2018lfu}. A summary of the current limits on different corresponding parameters at the LHC can be found on the Ref.\cite{AguilarSaavedra:2018nen} giving the limits on the Wilson Coefficients of the top involved four-fermion operators including two quark and two leptons at $\mathcal{O}(0.1)$ assuming $\Lambda = 1$ TeV. Such studies have also been carried out in the $e^-e^+$ colliders \cite{BarShalom:1999iy, Abramowicz:2018rjq,Escamilla:2017pvd,Durieux:2017gxd,Shi:2019epw,Durieux:2018tev} yielding the most stringent limits on some Wilson Coefficient such as $C_{lq}^{(3)ee33}$ at $\mathcal{O}(0.01)$ assuming a relatively larger $\Lambda = 10$ TeV \cite{Escamilla:2017pvd}. This is obtained by assuming the experimental uncertainty to be as high as 30$\%$. While at the LHeC, most efforts have been put at dimension four FCNC operators \cite{Behera:2018ryv, Cakir:2018ruj, TurkCakir:2017rvu, Wang:2017pdg, Sun:2016kek, Liu:2015kkp}. The exact process $e^{-} p \to e^{-} t$ has been considered at a recent paper \cite{Behera:2018ryv} while focusing on anomalous $Ztq$ couplings. Apart from these, a study at the LHeC about the dimension six operator containing the Higgs boson has been carried out in Ref. \cite{Hesari:2018ssq}. Ref.\cite{Duarte:2014zea,Alcaide:2019pnf} studies a similar quark-quark-lepton-lepton ($QQLL$) operator for different generation which does not contain top quark but include a majorana neutrinos. A similar research considering neutrino in the final states rather than the electron, while extra $tbW$ interaction has been considered with a focus on non flavour changing operators has been shown in Ref. \cite{Sarmiento-Alvarado:2014eha}. 

In this paper, we focus on the operators in the basis of dimension six SMEFT which can generate a single top production $e^{-} p \to e^{-} t$ process at the LHeC, i.e. four-fermion operators at 1113 generation. We select the scale factor $\Lambda$ in the Wilson coefficient to be 1 TeV for comparison to the other researches. Particularly, we reconstruct the top in the final states using the leptonic channel of its decay, $b$, $e^{-}$ and $\slashed{E}^{miss}_T$ regrading the current capability of the full detector level simulation of LHeC. By putting cuts on the reconstructed top mass and the transverse mass of W boson, a relatively high significance $S/\sqrt{B}$ is obtained. 

The paper is organized as follows: In Section \ref{sec:Theory}, we briefly review the dimension-six Standard Model Effective Field Theory and summarize the relevant operators for the process $e^{-} p \to e^{-} t$. Following this, we simulate the cross section of this signal process depending on the corresponding Wilson coefficient, analyze the main background at the LHeC, and put kinematical cuts on them to reduce the background and finally obtain the sensitivities in the Section \ref{sec:simulation}. In the end, we summarized the paper and give the conclusions in Section \ref{sec:con}.

\section{Dimension-six FCNC four-fermion contact operators in SMEFT}
\label{sec:Theory}

\vspace{1cm}

\subsection{Relevant Operators}

When physics beyond the SM is present at scales ($\Lambda$) larger than the electroweak scale, the SM can be extended into an effective field theory (EFT). The so called Standard Model Effective Field Theory (SMEFT) including the same symmetric group $SU(3)\times SU(2)\times U(1)$ of the SM, defined by a power counting expansion in the ratio of the scales, extends the SM with higher dimensional operators ${\cal Q}^{(d)}_i$ of mass dimension $d$. The lagrangian is expressed as
\begin{align} \nonumber
{\cal L}_\text{SMEFT} =
{\cal L}_\text{SM}^{(4)} + \sum_{k} \frac{C_{k}^{(5)} }{\Lambda} {\cal Q}_{k}^{(5)}
                         + \sum_{k} \frac{C_{k}^{(6)} }{\Lambda^2} {\cal Q}_{k}^{(6)} \\
+ {\cal O} \left( \frac{1}{\Lambda^{3}}\right)
\label{LBSM}
\end{align}
where ${\cal L}_\text{SM}^{(4)}$ is the SM lagrangian and $C_{k}^{(d)}$ stands for the corresponding dimensionless coupling constants (Wilson coefficients). The dimension-five terms vanish if baryon and lepton number conservation is imposed, and the dimension-six terms, which we concentrate on, are $\sum_{k} \frac{ C_{k}^{(6)} }{\Lambda^2} {\cal Q}_{k}^{(6)}$.

A complete set of all allowed dimension-six operators is actually quite large. However, not all operators, obeying the required symmetries of the SM lagrangian, are independent. They are related by the equations of motion and also by Fierz transformations. Therefore, the total number of operators can be reduced to a minimum set of independent ones. For the reason of simplify, they can be classified into three different groups: strong, electroweak and four-fermion operators \cite{Russell:2017cut}.

In our present paper, we are concentrating on the four-fermion operators where the complete non-redundant set of four-fermion operators are shown in \cite{Grzadkowski:2010es} and 
referenced in the so called "Warsaw basis". Here the operators should be supplemented
with generation indices of the fermion fields whenever necessary, e.g., $Q_{qq}^{(1)} \to Q_{qq}^{(1)prst}$, here $p,r,s,t= 1,2,3$ are generation indices. (1) or (3) are isospin indices, (1) and (8) are colour indices. Excluding the five B-violating operators, whose effects must be strongly suppressed to respect proton decay bounds, we have 59 independent operators. In fact, one can relax the flavour assumptions and allow all possible flavour combinations to be an independent operator. This increases the operator set to 2499 operators. The set of four-fermion operators can be further reduced when only specific processes are studied. For example, at electron proton (ep) colliders, ep collision can be used to study four-fermion interaction which involving two leptons and two quarks.
They may therefore sensitive to $Q_{lq}^{(1)}$, $Q_{lq}^{(3)}$, $Q_{\ell u}$, $Q_{eu}$, $Q_{lequ}^{(1)}$ and $Q_{lequ}^{(3)}$  which are shown as below
\begin{eqnarray} \nonumber
{\cal O}^{(1)prst}_{lq} &=& (\bar l_p\gamma_\mu l_r)  (\bar q_s\gamma^\mu q_t) ,\\\nonumber
{\cal O}^{(3)prst}_{lq} &=& (\bar l_p\gamma_\mu \tau^I l_r) (\bar q_s\gamma^\mu \tau^I q_t),\\\nonumber
{\cal O}^{prst}_{lu} 	&=& (\bar l_p\gamma_\mu l_r) (\bar u_s\gamma^\mu u_t) ,\\\nonumber
{\cal O}^{prst}_{eq}	&=& (\bar e_p\gamma_\mu e_r) (\bar q_s\gamma^\mu q_t) ,\\\nonumber
{\cal O}^{prst}_{eu} 	&=& (\bar e_p\gamma_\mu e_r) (\bar u_s\gamma^\mu u_t) ,\\\nonumber
{\cal O}^{(1)prst}_{lequ} &=& (\bar l_p e_r) \epsilon (\bar q_s u_t),\\
{\cal O}^{(3)prst}_{lequ} &=& (\bar l_p \sigma_{\mu\nu} e_r) \epsilon (\bar q_s \sigma^{\mu\nu} u_t).
\end{eqnarray}
The notation employed in this section is following that of Ref.~\cite{Grzadkowski:2010es} with flavour indices labelled by $prst$; left-handed fermion doublets denoted by $q$, $l$; right-handed fermion singlets by $u$, $d$, $e$; the antisymmetric $SU(2)$ tensor by $\varepsilon\equiv i\tau^2$; where $\tau^I$ are the Pauli matrices.

Further more, we are considering the single top production effects at ep colliders, so we should use a set of dimension six four fermions operators always involving at least one top quark and at least one electron. Our discussion exclusively concerns processes involving at least a top quark. Only operators involving such a particle are considered. Other operators affecting the considered processes are assumed to be well constrained by measurements in processes that do not involve top quarks. This assumption may not always be justified and explicit checks should be performed. 

\vspace{1cm}

\subsection{Current limits}

There are no direct limits of the top related flavour changing four-fermion operators so far. Anyhow, various studies about four-fermion operators can be applied to these operators through the measurement of CKM matrix elements. Indirect limits from low-energy observables, including B decays, dilepton production, electric dipole moments, CP asymmetries, proposed e− e+ colliders and future LHC upgrades such as HL-LHC as well as HE-LHC \cite{Cerri:2018ypt}, can apply to the operators mentioned above, here we only introduce the limits available at current experiments as discussed in the following subsections. We follow a similar procedure as in the Ref. \cite{AguilarSaavedra:2018nen}, while instead of assuming (33) element to be the only effective one, we assume a off diagonal (13) or (23) element to involve top flavour changing.

\paragraph{Limits from B physics}

Consider the charged-current $b\rightarrow c e_{i} \bar{\nu_{j}}$:
\begin{widetext}
\begin{equation}
H_{eff}^{b \rightarrow c e \bar{\nu_{j}}} = -\frac{2}{\nu^2} V_{cb}((\delta^{ij}+c_{V_{L}}^{ij})(\bar{c_L}\gamma^{\mu}b_L)(\bar{e_{L}^{i}}\gamma^{\mu}\nu_{L}^{j})+c_{S_{R}}^{ij}(\bar{c_{L}}b_{R})(\bar{e_{R}^{i}}\nu_{L}^{j})+h.c.)
 \label{eq:Heff}
\end{equation}
\end{widetext}
where $\nu \approx$ 246 GeV. The tree-level matching to the SMEFT is given by \cite{Cirigliano:2012ab, Aebischer:2015fzz}:
\begin{eqnarray}
 c_{V_{L}}^{ij}&=&-\frac{\nu^2}{\Lambda^2}(\frac{\Sigma_{k}V_{ck}C_{lq}^{(3)ij3k}}{V_{cb}})+\frac{\nu^2}{\Lambda^2}(\frac{\Sigma_{k}V_{ck}C_{\phi q}^{(3)k3}}{V_{cb}})\delta^{ij}\ \ \\
 c_{S_{R}}^{ij}&=&\frac{\nu^2}{\Lambda^2}(\frac{\Sigma_{k}V_{ck}C_{lq}^{(3)ij3k}}{V_{cb}}),
\end{eqnarray}
By assuming only the relevant operators in the paper non-vanishing, it can leads to:
\begin{eqnarray}
& c_{V_{L}}^{ij}=-\frac{\nu^2}{\Lambda^2}(\frac{V_{cd}C_{lq}^{(3)ij31}}{V_{cb}})
\end{eqnarray}
or 
\begin{eqnarray}
& c_{V_{L}}^{ij}=-\frac{\nu^2}{\Lambda^2}(\frac{V_{cs}C_{lq}^{(3)ij32}}{V_{cb}}).
\end{eqnarray}
For example, in the Ref.\cite{AguilarSaavedra:2018nen,Buttazzo:2017ixm}:
\begin{eqnarray}
& |1 + c_{V_{L}}^{\tau \tau}|^2=1.237\pm 0.053
\end{eqnarray}
Applying $c_{V_{L}}^{\tau \tau}=-\frac{\nu^2}{\Lambda^2}(\frac{V_{cb}C_{lq}^{(3)\tau \tau 33}}{V_{cb}})$, we obtained $C_{lq}^{(3)\tau \tau 33}=-1.85\pm 0.40$ which agreed on the result shown in the Ref. \cite{AguilarSaavedra:2018nen}. Thus, similar results can be obtained by translation through dividing the CKM matrix $\frac{V_{cd}}{V_{cb}}$ or $\frac{V_{cs}}{V_{cb}}$, with the translated results of Ref. \cite{Jung:2018lfu} shown in the Table \ref{tab:current_limits}.
\begin{table}
	\begin{center}	
		\begin{tabular}{|l|p{70mm}|}
			\hline
			$C_{lq}^{(1)}$, $C_{lq}^{(3)}$& 
			${C}_{lq}^{(3)ee31} \in$ [-0.10, 0.04] ([-0.024, 0.0093]) \cite{Jung:2018lfu},  ${C}_{lq}^{(1)ee31}+{C}_{lq}^{(3)ee31} \in$ [-0.059, 0.037] ([-0.014, 0.0085]) \cite{Greljo:2017vvb}, 
			\\ \hline
			$C_{eq}$& $C_{eq}^{ee31}$ $\in$ [-0.044, 0.051] ([-0.010, 0.012]) \cite{Greljo:2017vvb}   \\ \hline
			$C_{lu}$& \quad \quad \quad \quad \quad \quad No current limits  \\ \hline
			$C_{eu}$&  \quad \quad \quad \quad \quad \quad No current limits \\ \hline
			$C_{lequ}^{(1)}$&
			\quad \quad \quad \quad \quad \quad No current limits \\ \hline
			$C_{lequ}^{(3)}$&
			\quad \quad \quad \quad \quad \quad No current limits \\ \hline
		\end{tabular}	
	\end{center}
	\caption{Current limits on the relevant operators when $\Lambda$ = 1 TeV for 31 (32) or 13 (23) generations.}
	\label{tab:current_limits}
\end{table}

\paragraph{Limits from high-$p_{T}$ di-lepton searches}
The effective operators can be strictly limited by the high-energy tail of 2 $\rightarrow$ 2 scattering processes such as $qq\rightarrow ll$ in Ref. \cite{Greljo:2017vvb}. Such limits are only valid if only the maximal centre of mass energy $E_{max}$ is much lower than the mediated massive mediators $M_{NP}$ which have been integrated out. The limits on the flavour changing top operators can be obtained using the same method provided in the former subsection, as the results are shown in the Table \ref{tab:current_limits}.

\paragraph{Limits from electric dipole moments}
As electric dipole moments have been constrained very strictly, it can be sensitive to new physics including four-fermion operators. In the one loop level, the electric dipole moment can be induced by single top loop. Thus, at one loop level, only same flavour quarks contributes to this process, not applicable for the favour changing operators.

\vspace{1cm}

\section{Numerical Results}
\label{sec:simulation}

\subsection{The simulation at the LHeC}

For the simulation of the collider phenomenology, we use FeynRules\cite{Alloul:2013bka} to extract the Feynman
Rules from the Lagrangian. The model is generated into Universal FeynRules Output(UFO) files\cite{Degrande:2011ua} from Ref. \cite{AguilarSaavedra:2018nen} and then fed to the Monte Carlo event generator MadGraph@NLO~\cite{Alwall:2014hca} for the generation of event samples. In order to estimate the event rate at parton level for the signal and backgrounds, we apply the following basic pre-selections:
\begin{eqnarray}\label{generator-level-cuts}\nonumber
&& \rm p_T^{b,j,\ell} > 5\ GeV, \ \ |\eta^{b,j (\ell)}| < 8(5), \\
&& \rm \Delta R(k_1k_2) >0.4, \ \ k_1k_2=jj, j\ell, jb, bb, b\ell,
\end{eqnarray}
where $\rm \Delta R = \sqrt{\Delta \Phi^2 + \Delta \eta^2}$ is the separation in the rapidity($\rm \eta$)-azimuth($\rm \Phi$) plane, $\rm p_T^{j, b, \ell}$ are the transverse momentum of jets, b-jets and leptons. $\eta$ is the pseudorapidity for the corresponding particles, while the cuts put here are only to create initial samples as we will pass the samples to more delicate cuts for the LHeC detectors introduced below. The cuts are defined in the lab frame. We pass the generated parton level events on to PYTHIA6.4~\cite{Sjostrand:2006za} which handles the initial and final state parton shower, hadronization, heavy hadron decays, etc. Delphes3.4.1~\cite{Ovyn:2009tx} is used for detector simulation with HEPMC \cite{Dobbs:2001ck} file as the input. The detector is assumed to have a cylindrical geometry comprising a central tracker followed by an electromagnetic and a hadronic calorimeter. The forward and backward regions are also covered by a tracker, an electromagnetic and a hadronic calorimeter. The angular acceptance for charged tracks in the pseudorapidity range of$-4.3 < \eta < 4.9$ and the detector performance in terms of momentum and energy resolution of electrons, muons, and jets, are based on the LHeC detector design~\cite{AbelleiraFernandez:2012cc,LHeCDetector} (Details here are taken from our former paper\cite{Azuelos:2017dqw}). For our simulation, a modified pythia version tuned for the ep colliders and the delphes card files for the LHeC detector configurations\cite{communication} are used. We use NN23LO1\cite{Deans:2013mha}\cite{Ball:2012cx} parton distribution functions for all event generations. 
The factorisation and renormalisation scales for both the signal and the background simulations are done with the default MadGraph5 dynamic scales. Antikt algorithm is adopted by jet definition. The other parameters like w boson mass is chosen to be $M_{W} =$ 80.385 GeV \cite{Patrignani:2016xqp} and top quark as $M_{t} = 172.32$ GeV. The collision energy is chosen to be 60 GeV electron beam and 7 TeV proton beam as proposed.

\subsection{The signal production at the LHeC}

\begin{figure}[htp]
\centering
   \includegraphics[height=4.3cm,width=7.4cm]{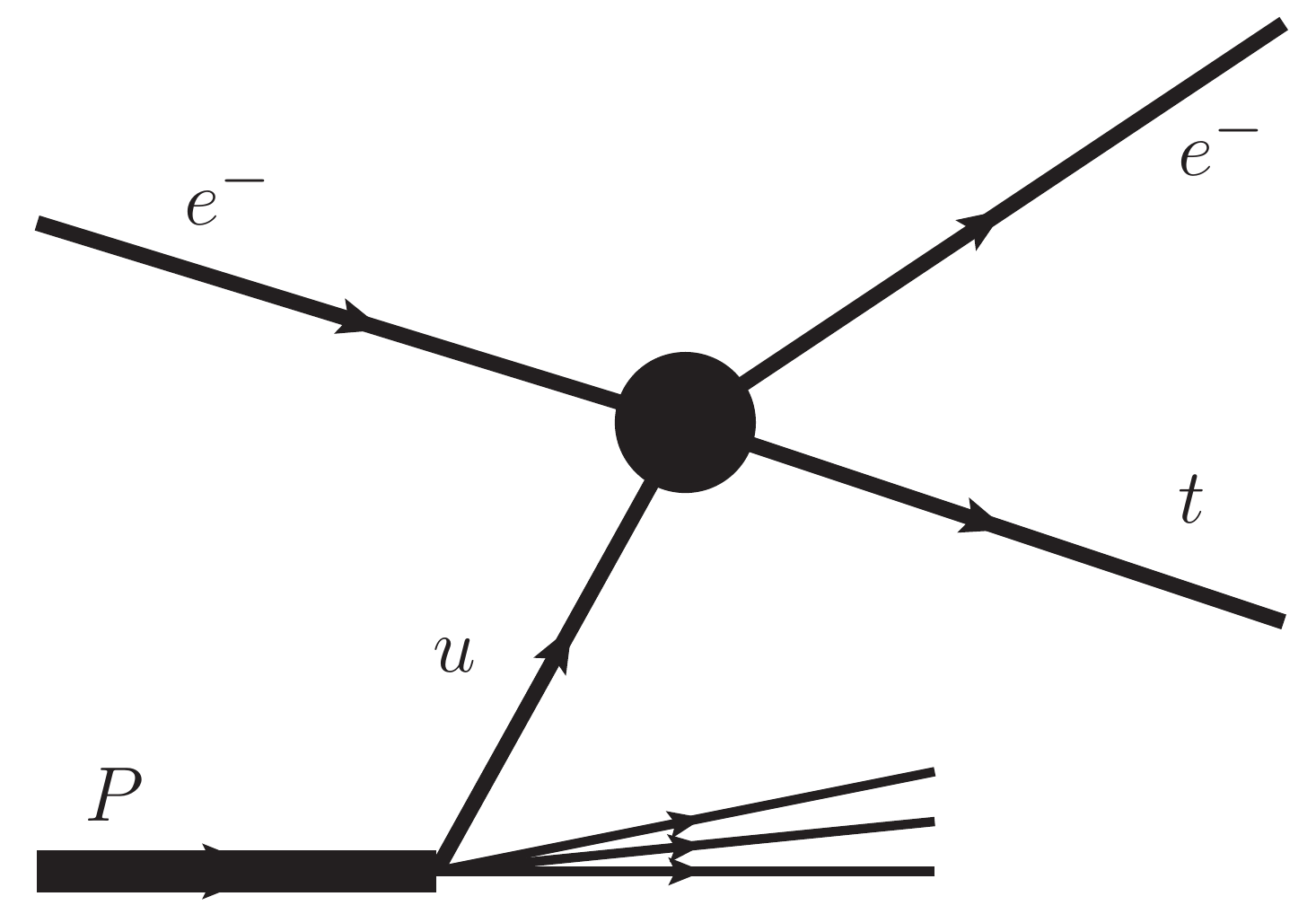}
   \caption{ \normalsize Signal Production through ep collision.}
   \label{fig:crossX}
\end{figure}

We start with
\begin{eqnarray}
 e^- (p_1) + u (p_2) \to e^- (p_3) + t (p_4).
\end{eqnarray}
The process involving the second generation can be calculated analogously, and the contribution of the third generation is negligible. The Feynman diagram for the process is shown in Fig.\ref{fig:crossX}.

After Fierz transformation the part of Lagrangian containing the above four-Fermi operators can be parameterized as follows \cite{BarShalom:1999iy}:
\begin{widetext}
\begin{eqnarray}
\begin{split}
{\cal L}_\text{tuee} = \frac{1}{\Lambda^2} &\sum_{i,j=L,R}[ V_{ij} (\bar e \gamma_\mu P_i e)  (\bar t \gamma^\mu P_j u)+ S_{ij} (\bar e P_i e)(\bar t P_j u) + T_{ij} (\bar{e} \sigma_{\mu \nu} P_i e)(\bar{t} \sigma^{\mu\nu} P_j u)]
\end{split}
\end{eqnarray}
\end{widetext}
where $P_{L,R} = (1 \mp \gamma_5)/2$, $\sigma^{\mu\nu} = \frac{i}{4} [\gamma^\mu \gamma^\nu-\gamma^\nu \gamma^\mu]$ and express these vector-like ($V_{ij}$), scalar-like ($S_{ij}$) and tensor-like ($T_{ij}$) couplings in terms of the coefficients of the seven four-Fermi operators we mentioned above. Applying Fierz transformation, we get the coefficient to the following terms
\begin{eqnarray} \nonumber
V_{LL}&=& C_{\ell q}^{(1)} - C_{\ell q}^{(3)}  + C_{\ell q}^{(1)*} - C_{\ell q}^{(3)*} , \\\nonumber
V_{LR}&=& C_{\ell u} + C^{*}_{\ell u}, \ \ V_{RL} = C_{qe} + C^{*}_{qe}, \  \\\nonumber
V_{RR}&=& C_{eu} + C^{*}_{eu},  \\\nonumber
S_{RR}&=& - ( C^{(1)}_{\ell e qu} + C^{(1)*}_{\ell e qu} ), \\\nonumber
S_{LL}&=& S_{LR}=S_{RL}=0 , \\\nonumber
T_{RR}&=&  \frac{1}{4} ( C^{(3)}_{\ell equ } + C^{(3)*}_{\ell equ } ) , \\ T_{LL}&=&T_{LR}=T_{RL}=0,
\end{eqnarray}
The corresponding amplitude can be written as
\begin{eqnarray}\nonumber
|{\cal M}^V_{LL}|^{2}
&=& \frac{V^2_{LL}}{\Lambda^4} 4 \hat{s}(\hat{s}-m_t^2),
|{\cal M}^V_{LR}|^{2}
= \frac{V^2_{LR}}{\Lambda^4} 4 \hat{u}(\hat{u}-m_t^2)   \\\nonumber
|{\cal M}^V_{RL}|^{2}
&=& \frac{V^2_{RL}}{\Lambda^4} 4 \hat{u}(\hat{u}-m_t^2),
|{\cal M}^V_{RR}|^{2}
= \frac{V^2_{RR}}{\Lambda^4} 4 \hat{s}(\hat{s}-m_t^2)   \\\nonumber
|{\cal M}^S_{RR}|^{2}
&=& \frac{S^2_{RR}}{\Lambda^4} \ \hat{t}(\hat{t}-m_t^2)   \\
|{\cal M}^T_{RR}|^{2}
&=& \frac{T^2_{RR}}{\Lambda^4} [ 2 \hat{s}(\hat{s}-m_t^2) - \hat{t}(\hat{t}-m_t^2) + 2 \hat{u}(\hat{u}-m_t^2) ] \ \ \ \ \ \
\end{eqnarray}
where $\hat{s}=(p_1+p_2)^2=(p_3+p_4)^2$, $\hat{t}=(p_1-p_2)^2$, $\hat{u}=(p_1-p_3)^2$.
So that we have
\begin{eqnarray}
\frac{ d\hat\sigma }{d\Omega} = \frac{1}{64\pi^2 \hat{s}} \frac{\hat{s}-m^2_t}{\hat{s}} \overline{{\cal |M|}^2}
\end{eqnarray}
with
\begin{eqnarray}
 \overline{ |{\cal M}|^2 } = \frac{1}{4} \sum_{ij=L,R} ( |{\cal M}^V_{ij}|^{2} + |{\cal M}^S_{ij}|^{2} +  |{\cal M}^T_{ij}|^{2}  )
\end{eqnarray}
The total cross section can be written as
\begin{eqnarray}
 \sigma = \int^1_{x^{min}_p} dx_p \sum_{q,\bar{q}} f_{q/p}(x_p, \mu^2_f) \left( \frac{d\hat\sigma}{d\Omega} \right) d\Omega
\end{eqnarray}
with $x^{min}_p = m^2_t /s $ and $\hat{s} = x_p s$.

As for the decay of the top quark in the LHeC, as the Standard Model $twb$ coupling is much bigger than the constrained FCNC couplings, the branching ratio for top decays to FCNC processes are considered negligible for this paper, which is also justified in several references \cite{Degrande:2013md, Sarmiento-Alvarado:2014eha}.  

The dependence on all independent parameters is just simply:
\begin{widetext}
\begin{multline}
	\sigma(e^{-} p \rightarrow t e^{-})\simeq 1.63 \cdot  ({V_{LL}}^2+{V_{RR}}^2) + 0.576 \cdot ({V_{LR}}^2 +  {V_{RL}}^2)
	+0.145 \cdot {S_{RR}}^2 +246 \cdot {T_{RR}}^2  [\text{pb}]
	\label{equ:all_dependence}
\end{multline}
\end{widetext}
Or in a basis of $C$ in the Wilson coefficient:
\begin{widetext}
\begin{eqnarray}\nonumber
\sigma(e^{-} p \to t e^{-}) &\simeq & 1.63 \cdot  （|({C_{lq}^{(1)ee31}-C_{lq}^{(3)ee31}})|^2+ |C_{eu}^{ee31}|^2) 
+0.576 \cdot (|{C_{eq}^{ee31}}|^2 + |{C_{lu}^{ee31}}|^2) \\
&+& 0.145 \cdot (|{C_{lequ}^{(1)ee13}}|^2 + |{C_{lequ}^{(1)ee31}}|^2)
+15.6 \cdot(|{C_{lequ}^{(3)ee13}}|^2 + |{C_{lequ}^{(3)ee31}}|^2) [\text{pb}] \ \ \ \
\end{eqnarray}	
\end{widetext}
Here we use $\Lambda$ = 1 TeV as the default set.
Use analogous methods for other partons, we can get the dependence for different generations. In the calculations following, we take the condition that only $C_{lq}^{(1)ee31}$ exists and equal to 1 as an example, and times the ratio of cross section of other parameters to get the corresponding effective cross sections for other parameters.

\subsection{The backgrounds at the LHeC}

The dominant SM background will be inclusive processes which contain the corresponding final states: 
$\ell^+ + \ell^- + \slashed{E}^{miss}_T +  b/\bar{b}/j$. At the LHeC, there are three dominant background: Vector boson fusion processes at the LHeC can introduce the same final states and the b jet is coming from the scattering of the initial proton beam; Inclusive top quark production processes at the LHeC can lead to similar signatures from its top decays， it is actually quite large as its cross section reaches 3.5 pb at the our setup of LHeC; Processes including $Z$ bosons can have rich leptonic and hardonic final states, thus can be sources of background. The above background can be summarized as:

a) $e^- p \to l^+ + e^- + \slashed{E}^{miss}_T +  b/\bar{b}/j$
or $e^- p \to l^- + e^- + \slashed{E}^{miss}_T +  b/\bar{b}/j$

b) $e^- p \to  \bar{t} + all$ 
or $e^- p \to  \bar{t} + all + all$
or $e^- p \to  \bar{t} + t + all + all$

c) $e^- p \to  e^- + z + z + j$
or $e^- p \to  \slashed{E}^{miss}_T + z + z + j$
or $e^- p \to  \slashed{E}^{miss}_T + z + j$
or $e^- p \to  e^{-} + z + j$

Note all inclusive processes containing more jets for the mentioned background are negligible.

In the next subsection, we will show that all the above background can be effectively removed using several kinematical cuts listed follows. That is mainly due to the requirement of the final states to from single top decays. The rich sources of the leptonic final states in the background makes it possible to distinguish signal from background by requiring exact number of leptons to be the decay products of single top and W boson.

\subsection{Kinematical Cuts}
\label{sec:cuts}

In order to distinguish signal from background, we put kinematical cuts to both of the signal and background:
\begin{itemize}
	\item Cut I: Requiring exactly one electron, one positron, one $\slashed{E}^{miss}_T$ and only one b-tagged jet with minus charge as the $b$ jet is coming from the massive particle $t$.
	\item Cut II: Based on Cut I, requiring the transverse mass $|M(e^{+}, \slashed{E}^{miss}_T )-M_{W}| < 10$ GeV.
	\item Cut III: Based on Cut II, requiring the invariant mass $|M(e^{+}, \slashed{E}^{miss}_T, b) - M_{t}| < 40$ GeV.
\end{itemize}

\begin{figure}[htp]
\centering
   \includegraphics[scale=0.5]{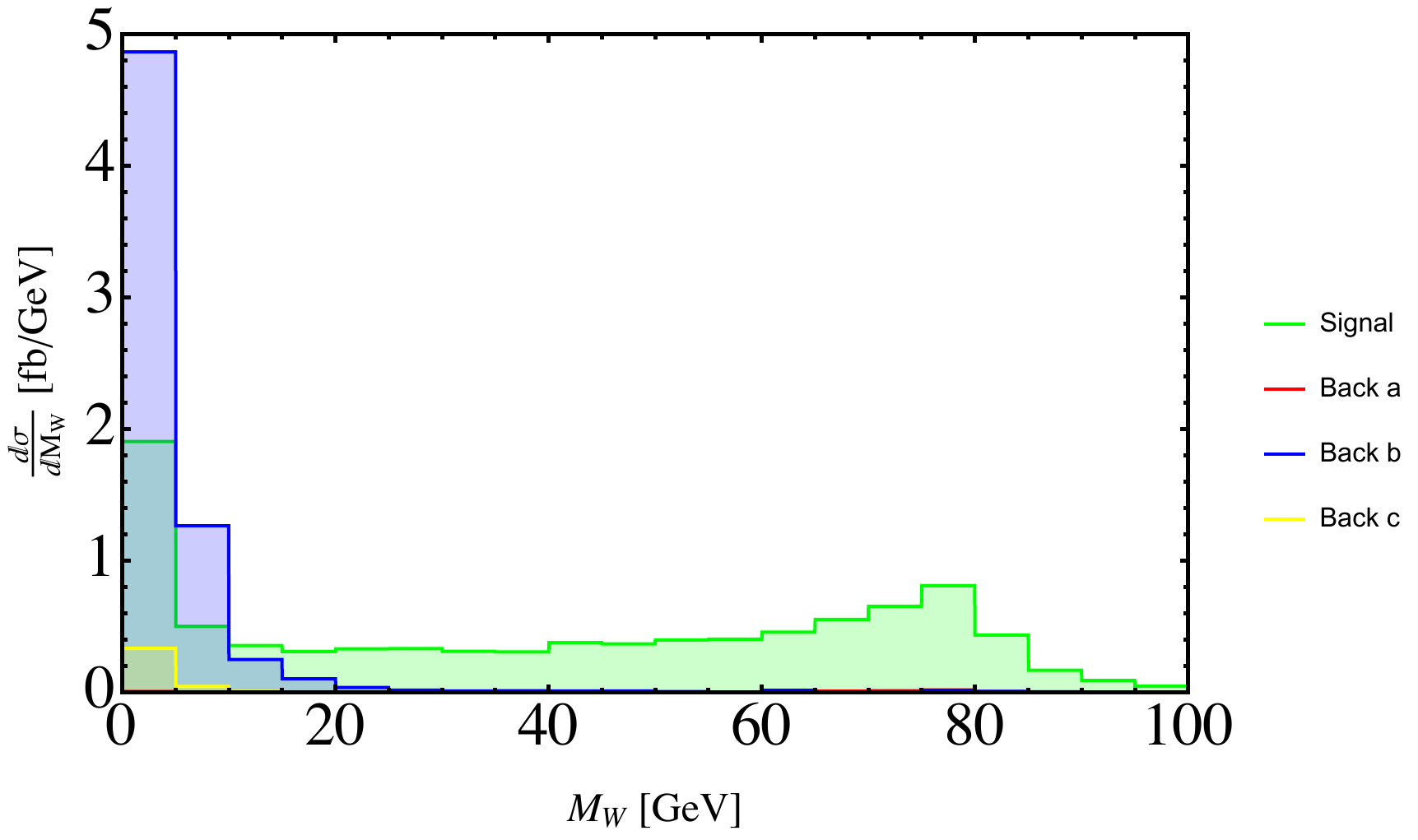}
   \caption{Reconstructed $W$ boson transverse mass from $e^{+}$ and $\slashed{E}^{miss}_T$ after Cut I. The Green bar presents the signal, while Red, Blue and Yellow represents for background a, b and c respectively. The bin size is 5 GeV.}
   \label{fig:wmass}
\end{figure}

\begin{figure}[htp]
\centering
   \includegraphics[scale=0.5]{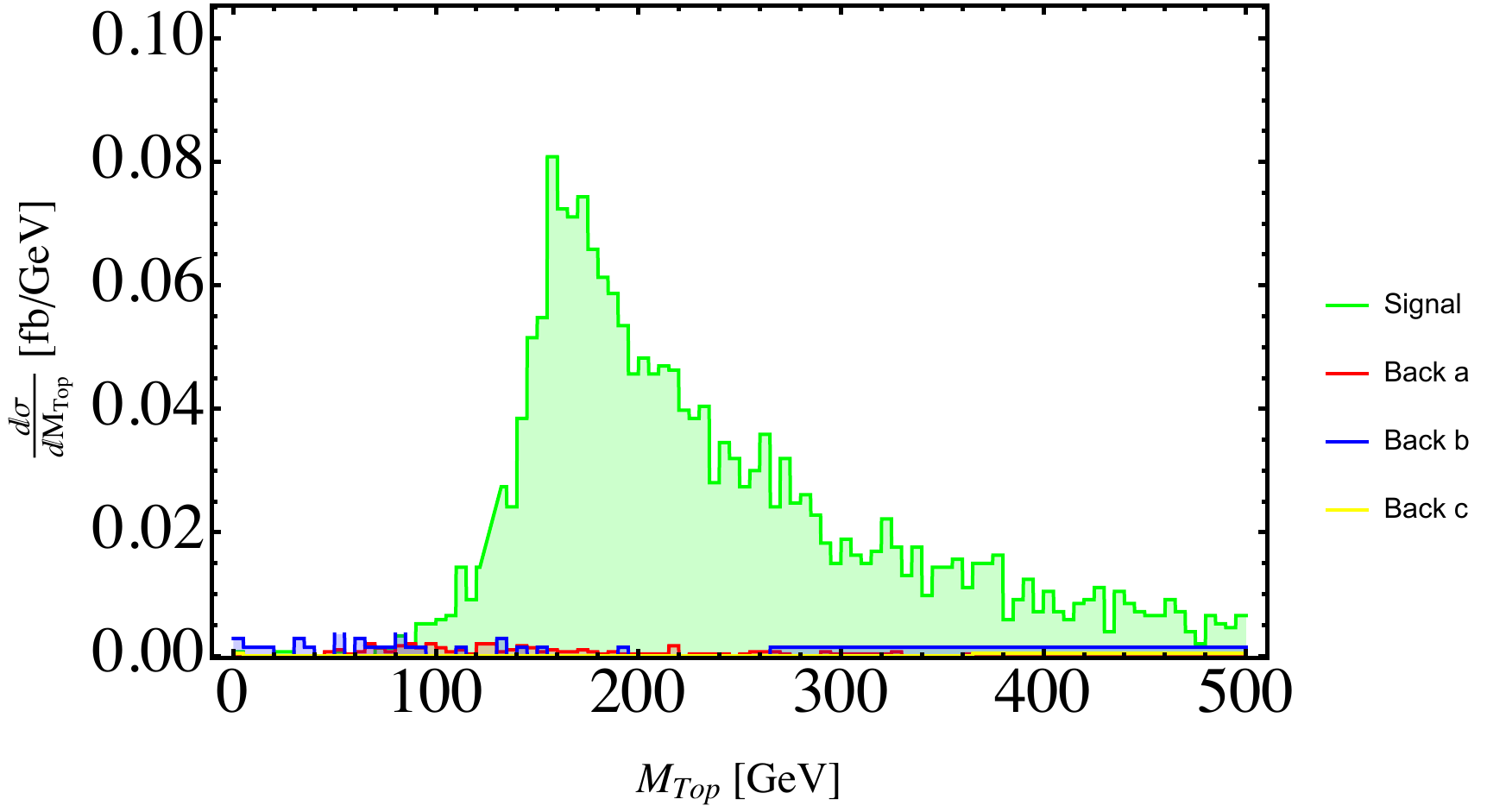}
   \caption{Reconstructed Top mass from $W$ and $b$ tagged jets after Cut II. The Green bar presents the signal, while Red, Blue and Yellow represents for background a, b and c respectively. The bin size is 5 GeV.}
   \label{fig:topmass}
\end{figure}

\begin{figure}[htp]
\centering
   \includegraphics[height=4cm,width=4cm]{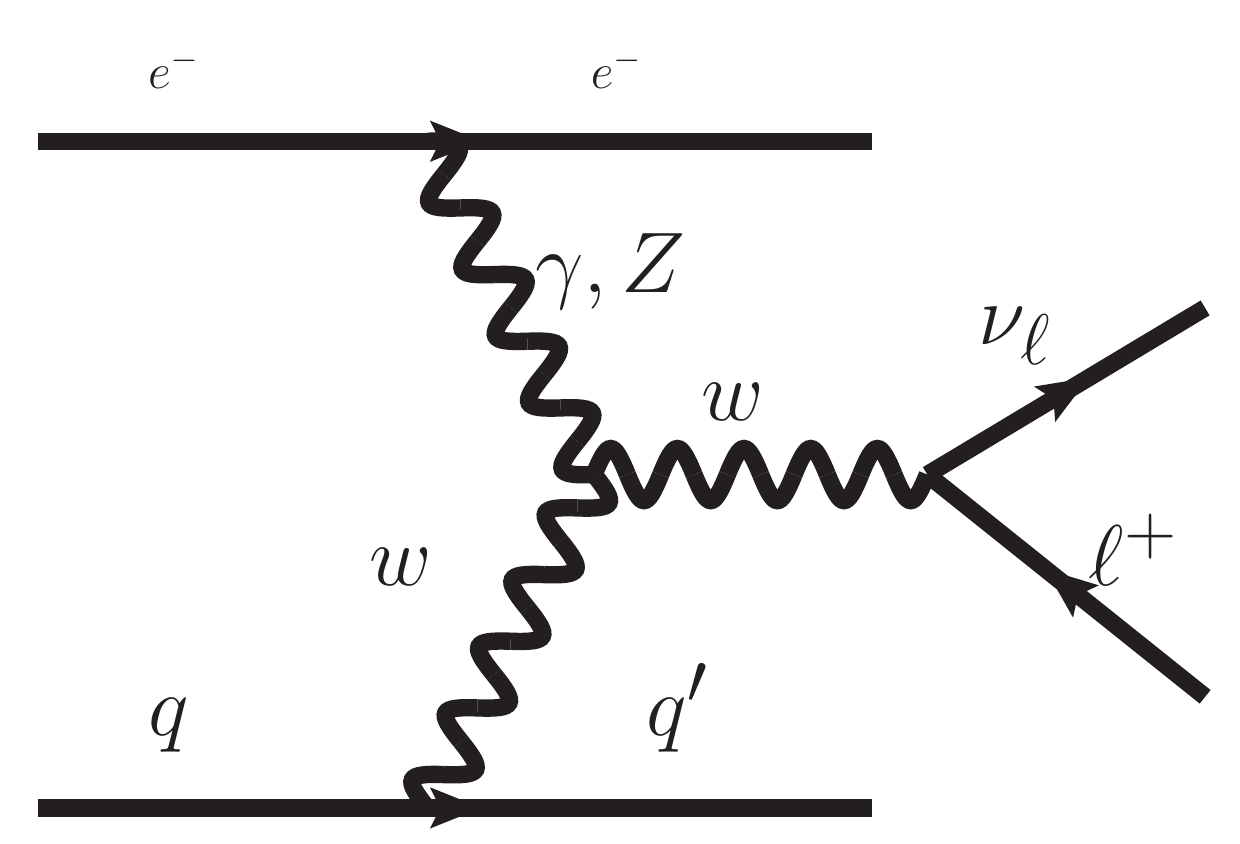}
   \caption{ \normalsize One example of background process $e^{-} p \rightarrow l^{+} + e^{-} + \nu_{l} + b $ at LHeC which contains W bosons as mediator for $l^{+}$ and $\nu_{l}$.}
   \label{fig:backgound}
\end{figure}

Note here we only considered the $e^{+}+\slashed{E}^{miss}_T+b $ final states of top for simplicity.  The antimuon final states can also be considered at future detector simulation.
We produced 500 thousand events for each signal and background, the cross section and the efficiency which is defined as $\frac{N(\text{after cuts})}{N(\text{initial events})}$ for all of signal and background is shown in the Table \ref{tab:cutflow}.
\begin{table*}
	\begin{center}	
		\begin{tabular}{|c|c|c|c|c|}
			\hline
			processes & signal & background a & background b &  background c  \\ \hline
			$\sigma$(fb) & 1630 & 818 & 3520  &  1016\\ \hline
			Cut I efficiency &  2.83e-02 & 8.06e-04 & 9.60e-03  & 2.04e-03  \\ \hline
			Cut II efficiency & 6.28e-03 & 2.88e-04 & 5.85e-05  & 6.00e-06  \\ \hline
			Cut III efficiency &  2.46e-03 & 1.08e-04 & 1.61e-05  & 2.00e-6  \\ \hline
			effective events (2 $ab^{-1}$) & 8020 & 177 & 113 & 4  \\ \hline
		\end{tabular}	
	\end{center}
	\caption{Cut flows for the signal and background process applying the kinematical cuts mentioned in the section \ref{sec:cuts} at the LHeC. The efficiency is defined as $N_{\text{after cuts}}/N_{\text{initial events}}$. }
	\label{tab:cutflow}
\end{table*}
From Table \ref{tab:cutflow}, we can see Cut I is quite efficient at distinguish between the signal and background even though only 13.3$\%$ \cite{Patrignani:2016xqp} of the top can have $e^{+}, \slashed{E}^{miss}_T$ at the final states. Cut II is extremely good at distinguish background containing top quark and $Z$ boson. For background b, this is due to the fact that the missing energy and positron in background b which mostly contain multiple tops can have different sources, thus it is not likely for them to reconstruct the $W$ boson mass. As for background c, this is quite straight forward as it is rare for $Z$ boson to decay into $W$ boson. For background a, however there are processes which contain $W\rightarrow e^{+} \nu_{e}$ which is shown in figure \ref{fig:backgound} makes it difficult to separate the background from the signal. Luckily, this background has been already reduced by the only one b-tagged requirement of Cut I. That is due to the fact that other than produced by the parenting top quark, the b-tagged jets in the background a are produced by the decay of gluons which are more likely to produce more than one b-tagged jets which have minus sign. As we already get big enough $S/\sqrt{B}$ from the above Cuts, the Cuts III we put to reconstruct the top mass is quite loose and not very effective. This selection of cuts are also justified in the Figure \ref{fig:wmass} and  \ref{fig:topmass}. Especially in figure \ref{fig:topmass}, it is shown that we are able to get a good distribution of reconstructed top mass with much higher efficiency comparing to the background which indicates that the Cuts we applied are efficient at distinguishing the signal from background.

\subsection{Sensitivity limits}
\label{sec:sensitivity}

With the cross section and efficiencies calculated above, we then apply a Goodness-of-fit with the method of Least Squares ($\chi^2$) analysis requiring $\chi^2 = (N_{tot}-N_B)^2/N_B > 3.84$ at 95 $\%$ C.L. \cite{Patrignani:2016xqp} at the LHeC with 2 $\text{ab}^{-1}$ luminosity. Thus the corresponding limits for different parameters in 31 and 13 generations assuming only one parameter existing are calculated by times its own cross sections and are shown in the table \ref{tab:limits}.
\begin{table}
	\begin{center}	
		\begin{tabular}{|c|c|c|c|}
			\hline
			Parameters & $|C_{lq}^{(1)ee31}|$&$|C_{lq}^{(3)ee31}|$ & $|C_{eu}^{ee31}|$    \\ \hline
			Limits & 0.0647 (0.367) & 0.0647 (0.367) & 0.0647 (0.367)    \\ \hline
			Parameters & $|C_{eq}^{ee31}|$ & $|C_{lu}^{ee31}|$ & $|C_{lequ}^{(1)ee13}|$     \\ \hline
			Limits  &  0.109(0.617) & 0.109(0.617) & 0.217 (1.23)   \\ \hline
			Parameters & $|C_{lequ}^{(1)ee31}|$ & $|C_{lequ}^{(3)ee13}|$ & $|C_{lequ}^{(3)ee31}|$  \\ \hline
			Limits & 0.217 (1.23) & 0.0209 (0.102) & 0.0209 (0.119)  \\ \hline
		\end{tabular}	
	\end{center}
	\caption{limits for the corresponding coefficient for the 31 and 13 (32 and 23) generation from $e^{+} + \nu_e $ channel at LHeC. The limits are calculated at 95 $\%$ C.L. for 2 $\text{ab}^{-1}$ LHeC. }
	\label{tab:limits}
\end{table}
Or summaried in the basis of independent parameter:
\begin{itemize}
	\item $\sqrt{({V_{LL}^{31}}^2+{V_{RR}^{31}}^2)} < 6.47 \times 10^{-2}$ 
	\item $\sqrt{({V_{LR}^{31}}^2 +  {V_{RL}^{31}}^2)} <$ 1.09 $\times 10^{-1}$
	\item $\sqrt{{S_{RR}^{31}}^2} <$ 2.17 $\times 10^{-1}$
	\item $\sqrt{{T_{RR}^{31}}^2} <$ 5.27 $\times 10^{-3}$.
\end{itemize}
The limits for 32 and 23 generations can be obtained by times a simple 32.2 factor, due to the different
content of the u-quark and c-quark in the LHeC proton
beam.

\section{Conclusion}
\label{sec:con}

In this work we have considered the four-fermion top flavour changing neutral current operators in the basis of dimension six Standard Model Effective Field Theory. We studied the process $e^{-} p \to e^{-} t$ in particular at the proposed electron proton collider LHeC, using its signature to set limits for the involved operators in the dimension six SMEFT. We perform our simulation at the full detector level requiring $e^{-} + e^{+} + \slashed{E}^{miss}_T +  b$ selection specifically at the LHeC. The corresponding backgrounds are reduced by setting kinematical cuts requiring the reconstructed W boson transverse mass and top quark mass. The sensitivities at the 2 $\text{ab}^{-1}$ LHeC is calculated by using a goodness-of-fit with the method of least squares ($\chi^2$) analysis. We obtained upper limits for $|C_{lq}^{(1)ee31}|, |C_{lq}^{(3)ee31}|$, $|C_{eu}^{ee31}|$ $<$ 0.0647, $|C_{eq}^{ee31}|$, $|C_{lu}^{ee31}|$ $<$ 0.109, $|C_{lequ}^{(1)ee13}|$, $|C_{lequ}^{(1)ee31}|$ $<$ 0.217 and $|C_{lequ}^{(3)ee13}|$, $|C_{lequ}^{(3)ee31}|$ $<$ 0.0209 assuming only one operator exists in the dimension six SMEFT for couplings of quarks at the 1st and 3th generation. These limits are calculated at 95 $\%$ C.L. for 2 $\text{ab}^{-1}$ LHeC. The couplings between quarks in the 2nd and 3rd generation can be obtained similarly. 
However, this process has no sensitivities for the couplings between two 3rd generation as $e^{-} t \rightarrow e^- t$ is negligible. That is to say that we have already obtain a competitive limits in flavour changing four-fermion operators such as $C_{lq}^{(1,3)}$ and $C_{eq}$ at the LHeC using a simple process. We have also gain new limits in the operators including right handed singlet of quark such as $C_{lu}$, $C_{eu}$, $C_{lequ}^{(1,3)}$, which is not covered in the B flavour physics.

These direct limits are already comparable to the indirect limits such as precise measurements of $B$ physics at the LHC through the CKM matrix. $e^{+}, e^{-}$ colliders are expected to have even cleaner background, thus producing indirect upper limits which are about a magnitude smaller for the operators. 

\begin{acknowledgments}
W. L. is supported by the China Scholarship Council
(Grant CSC No. 2016 08060325). H. S. is supported by the National Natural Science Foundation of China (Grant No.11675033) and by the Fundamental Research Funds for the Central Universities (Grant No. DUT18LK27). 
\end{acknowledgments}


\bibliography{apssamp}


\end{document}